\documentclass[a4paper]{PoS}
\usepackage{natbib}

\title{Rapid Variability: What do we learn from correlated mm-/gamma-ray variability in jets ?}

\ShortTitle{S5 0716+714 : Rapid Variability study }

\author{\speaker{B. Rani}$^a$ \thanks{Member of the Max Planck Research School (IMPRS) for Astronomy and Astrophysics at the Universities of Bonn and Cologne.}, T. P.\ Krichbaum$^{a}$, L. Fuhrmann$^{a}$, B. Lott$^{b}$, M. B{\"o}ttcher$^{c}$ and J. Anton Zensus$^a$\\
{\bf On behalf of the {\it Fermi}/LAT Collaboration and F-GAMMA Team} \\
\llap{$^a$} Max Planck Institut f\"ur Radioastronomie, Auf dem H\"ugel 69, 53121 Bonn, Germany\\
\llap{$^b$} Universit{\'e} Bordeaux 1, CNRS/IN2p3, Centre d'Etudes Nucl{\'e}aires de Bordeaux Gradignan, 33175 Gradignan, France\\
\llap{$^c$} Astrophysical Institute, Department of Physics and Astronomy, Ohio University Athens, OH 45701, USA\\ 
E-mail: \email{brani@mpifr-bonn.mpg.de}}

\abstract{ 
Densely time sampled multi-frequency flux measurements
of the extreme BL Lac object S5 0716+714 over
the past three years allow us to study its broad-band variability,
and the detailed underlying physics, with emphasis on
the location and size of the emitting regions and the evolution with time.
We study the characteristics of some prominent
mm-/$\gamma$-ray flares in the context of the shock-in-jet model and investigate  the
location of the high energy emission region.  
The rapid rise and decay of the radio flares is in agreement with the formation of a shock 
and its evolution, if a geometrical variation is included in addition to intrinsic 
variations of the source.  We find evidence for a correlation between flux variations at $\gamma$-ray and 
radio frequencies. A two month time-delay between $\gamma$-ray and radio flares indicates a non-cospatial 
origin of $\gamma$-rays and radio flux variations in S5 0716+714. 
}

\FullConference{  Nuclei of Seyfert galaxies and QSOs - Central engine $\&$ conditions of star formation,\\
		November 6-8, 2012\\
		Max-Planck-Insitut f$\ddot{u}$r Radioastronomie (MPIfR), Bonn, Germany}

\begin{document}

\section{Introduction}
Blazars constitute a unique laboratory to probe jet formation and its relation to radio-to-$\gamma$-ray variability.
The current understanding implies  that relativistic shocks propagating down the jet provide a good description 
of a variety of observed phenomena in AGNs. To provide a framework for the observed flux variations, we tested 
the evolution of radio flares in context of the standard shock-in-jet model [1; 2]. 
A shock induced flare follows a particular trend in the turnover frequency -- turnover flux density ($S_{m}$ -- $\nu_{m}$)
diagram. The typical evolution of a flare in the $S_{m}$ -- $\nu_{m}$  
plane can be obtained by inspecting the 
$R$ (radius of jet)-dependence of the turnover frequency, $\nu_{m}$ and the turnover flux density, $S_{m}$ 
[see 3 for details]. During the first stage,  Compton losses are dominant and 
$\nu_{m}$ decreases with increasing radius, $R$, while $S_{m}$ increases. In the second stage, where 
synchrotron losses are the dominating energy loss mechanism, $\nu_{m}$ continues 
to decrease while $S_{m}$ remains almost constant. Both $S_{m}$ and $\nu_{m}$ 
decrease in the final, adiabatic stage. As a consequence, the $S_{m}$ -- $\nu_{m}$ diagram 
is a useful tool to explore the dominance of emission mechanisms during various phases of evolution of a flare.

We report here a radio to $\gamma$-ray variability study of the BL Lac object S5 0716+714. 
We tested the evolution of radio (cm and mm) flares in context of the standard shock-in-jet model following 
the $S_{m}$ -- $\nu_{m}$ diagram as discussed above.  
We also investigate the correlation of $\gamma$-ray activity with the emission at lower frequencies, 
focusing on the individual flares observed between August 2008 and January 2011.

\section{Multi-frequency light curves}
\begin{figure*}
   \centering
\includegraphics[scale=0.55,angle=-90]{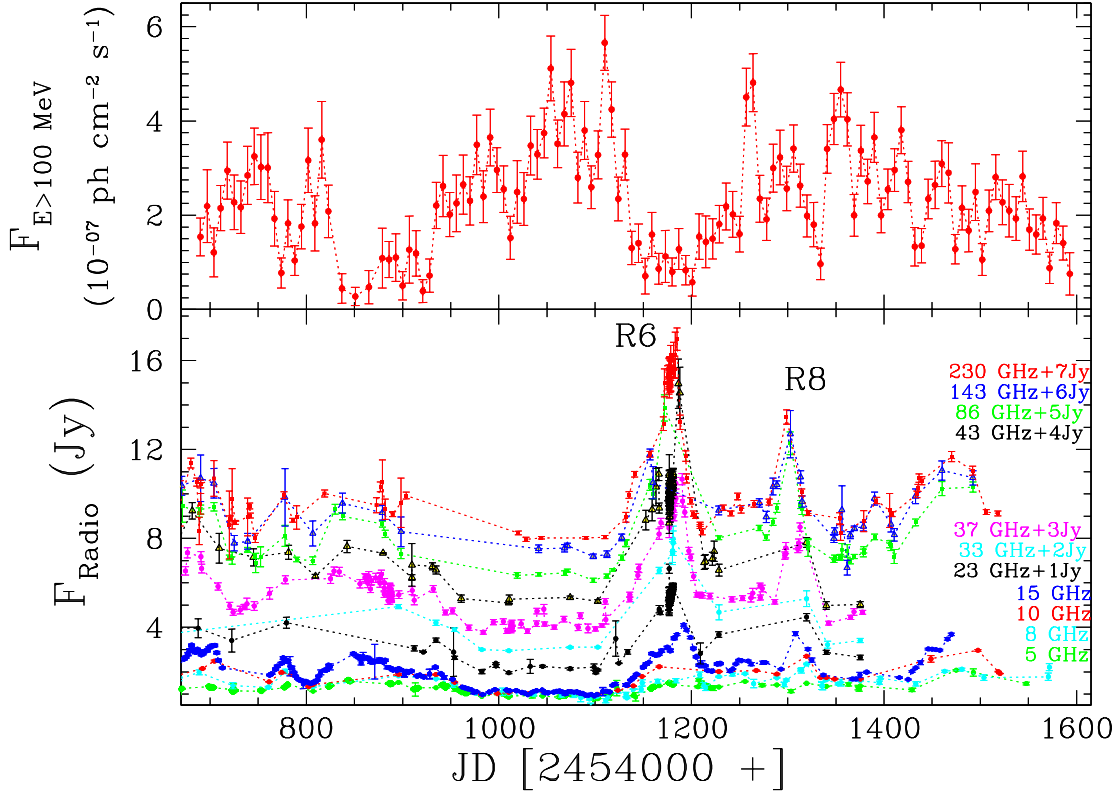} 
  \caption{Top : GeV light curve of S5 0716+714 during the 
first $\sim$3 years of the {\it Fermi}/LAT observations from 2008 August to 2011 January at E $>$ 100 MeV. 
Bottom : Radio frequency light curves of S5 0716+714 observed over the past $\sim$3 years. 
For clarity, the light curves at different frequencies are shown with arbitrary offsets 
(indicated by a "Frequency + x Jy" label).  The major radio flares are labeled as "R6" 
and "R8".              }
\label{plot_flx_rad}
    \end{figure*}

A broadband flux monitoring of S5 0716+714 was performed over a time period between April 2007 to January 2011. 
The multi-frequency observations comprise GeV monitoring by {\it Fermi}/LAT and radio monitoring by several 
ground based telescopes.  The details of observations and data reduction can be found in [4]. 
Fig. \ref{plot_flx_rad} shows the $\gamma$-ray and radio frequency light curves of the source. 
The top of the figure shows the weekly averaged $\gamma$-ray light curve 
integrated over the energy range 100 MeV to 300 GeV. 
The radio frequency light curves are shown in the bottom of the figure. The source exhibits significant flux variability 
both at $\gamma$-rays  and radio frequencies. Apparently, the two major radio flares (labeled as ``R6" and ``R8") are  observed 
after the major $\gamma$-ray flares.

\section{Evolution of radio flares in the shock-in-jet scenario}
\begin{figure*}
   \centering
\includegraphics[scale=0.25, angle=-90]{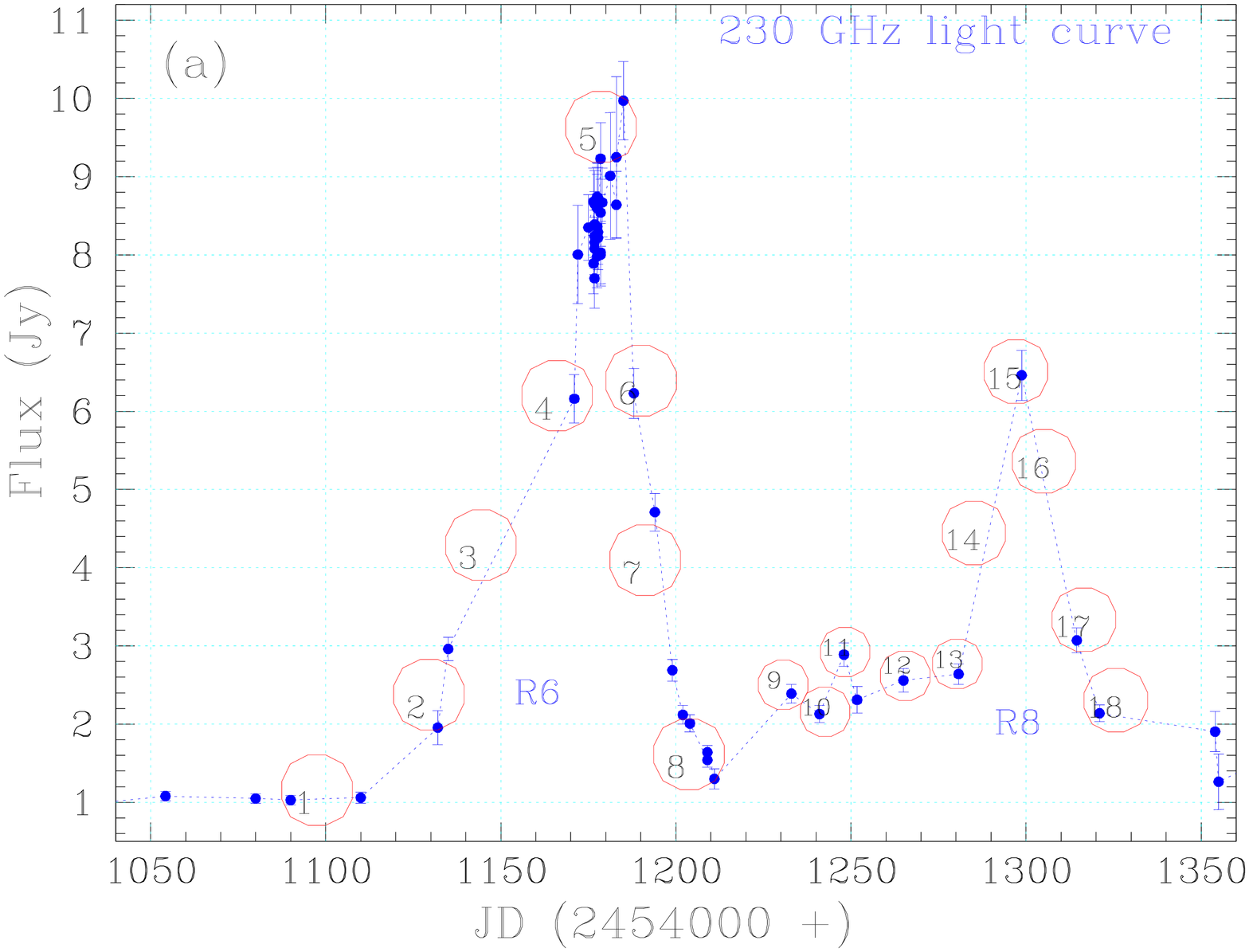}
\includegraphics[scale=0.25, angle=-90]{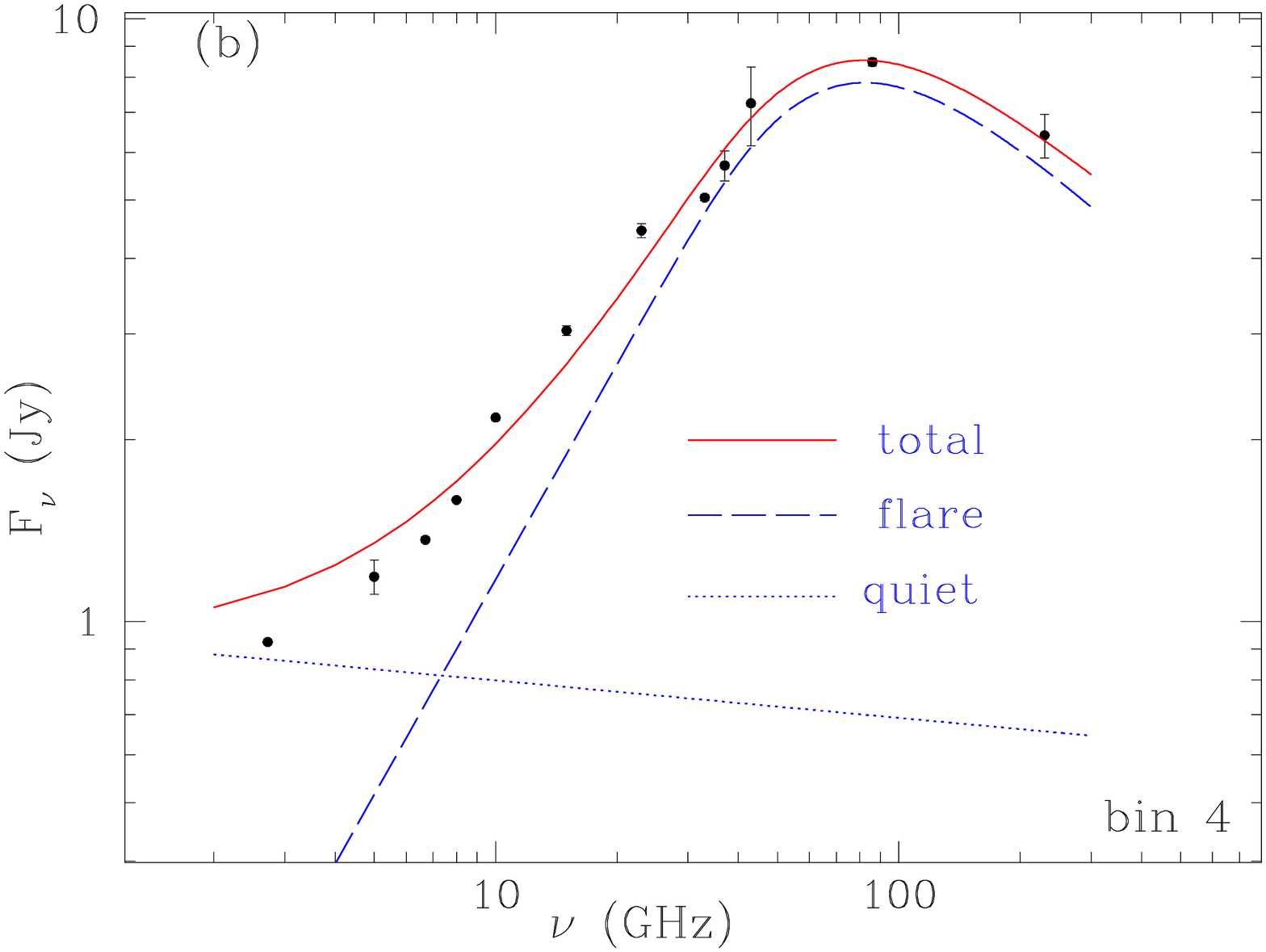}
\includegraphics[scale=0.25, angle=-90]{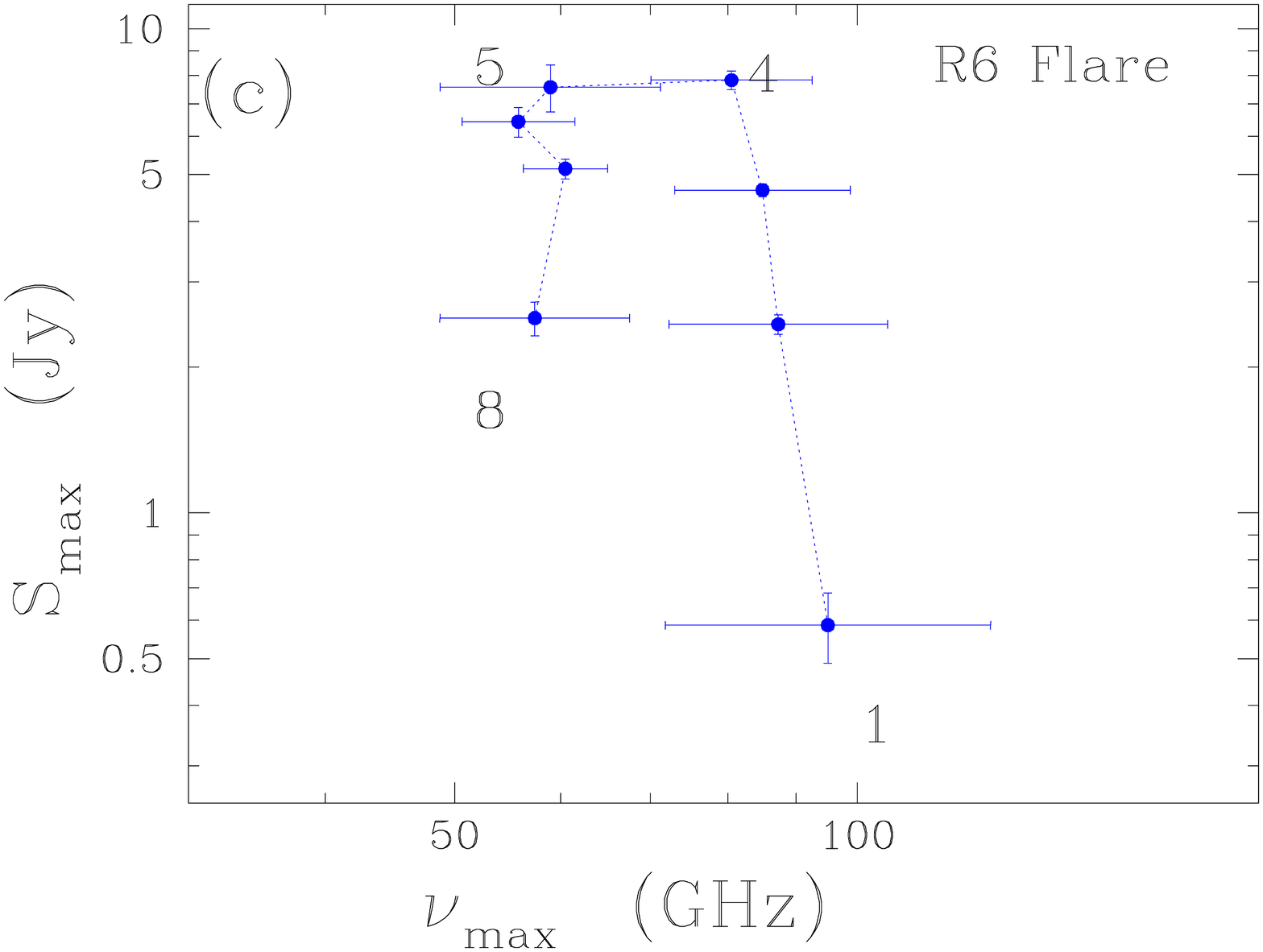}
\includegraphics[scale=0.25, angle=-90]{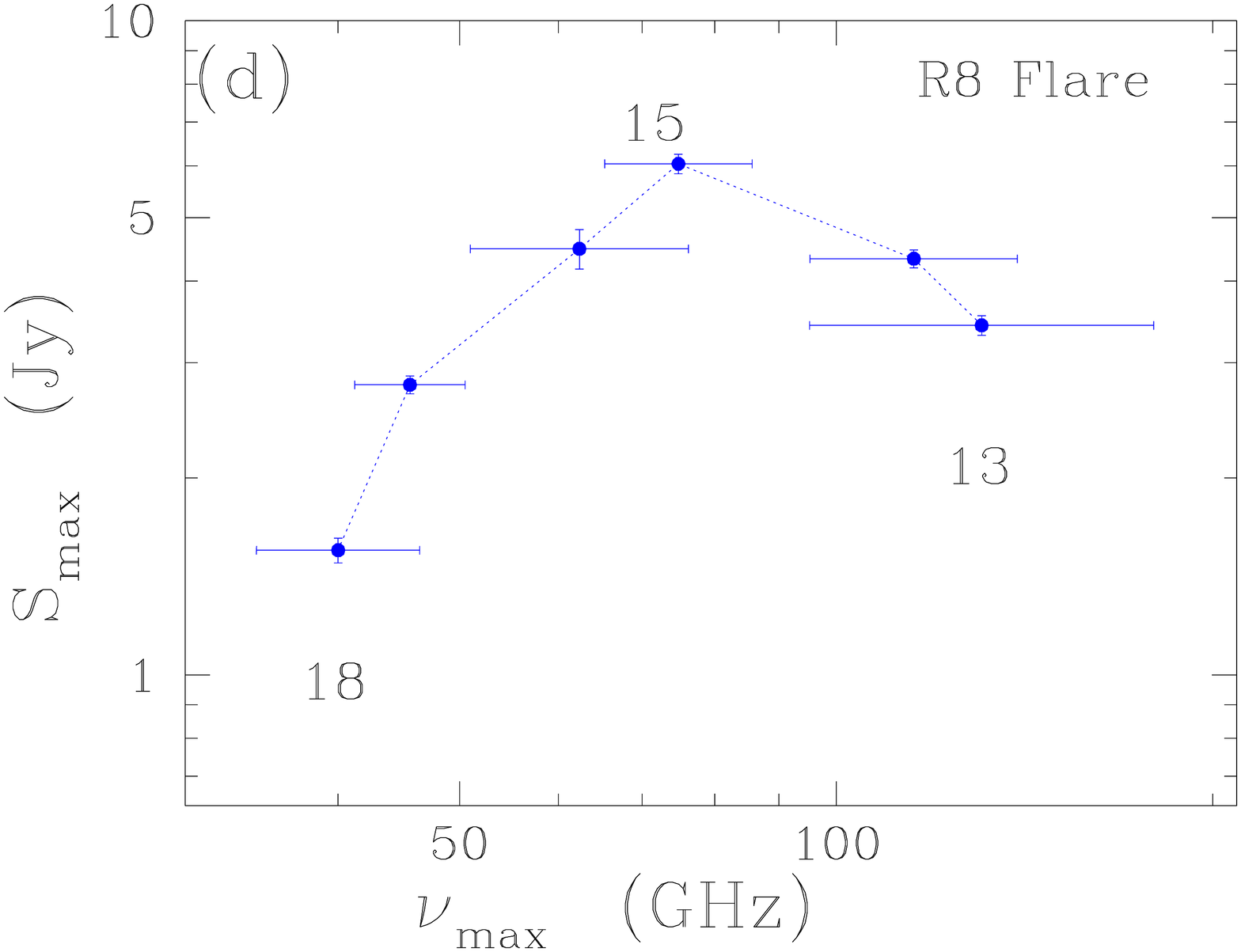}
   \caption{The evolution of the radio spectra: (a) 230~GHz light curve showing different periods 
over which the spectra are constructed.  (b) Results of a single component 
spectral fitting at time bin ``4", the dotted line corresponds to the quiescent spectrum, the dashed one to 
the flaring spectrum and the solid line to the total spectrum.  (c) $\&$ (d) The time evolution of $S_{max}$ vs $\nu_{max}$ 
for the R6 and R8 radio flares (see text for details).}
\label{radio_spectra} 
    \end{figure*}

In order to test the evolution of the two major radio flares in the context of a shock-in-jet model, we construct 
the quasi-simultaneous\footnote{time sampling $\Delta t = 5$~days} radio spectra 
over different time bins  as shown in Fig. \ref{radio_spectra} (a) [see 4 for details] 
using 2.7 to 230~GHz data. 
The observed radio spectrum is usually the superposition of emission from 
the two components : (i) a steady state (unperturbed region), and (ii) a flaring component resulting from 
the perturbed (shocked) regions of the jet.
The quiescent spectrum (Fig. \ref{radio_spectra} (b) (dotted curve)) is approximated using the lowest flux 
level during the course of our observations.
The quiescent spectrum  is described by a power law $F(\nu) = C_q (\nu/{\rm GHz})^{\alpha_{q}}$ with  
$C_q = (0.92 \pm 0.02)$~Jy and $\alpha_{q} = -(0.06\pm0.01)$. 
We subtract the contribution of the steady-state emission from the entire spectrum before modeling.

We fitted the flare component spectrum using a synchrotron self-absorbed model, which can be described 
as [see 3; 6 for details] : 
\begin{equation}
S_\nu=S_m\left(\frac{\nu}{\nu_m}\right)^{\alpha_t}\frac{1-\exp{\left(-\tau_m\left(\nu/\nu_m\right)^{\alpha_0-\alpha_t}\right)}}{1-\exp{(-\tau_m)}},
\label{snu}
\end{equation}
where $\tau_m\approx3/2\left(\sqrt{1-\frac{8\alpha_0}{3\alpha_t}}-1\right)$ is the optical depth at the turnover 
frequency, $S_m$ is the turnover flux density, $\nu_m$ is the turnover frequency and $\alpha_t$ and 
$\alpha_0$ are the spectral indices for the optically thick and optically thin parts of the spectrum, 
respectively ($S \sim \nu^{\alpha}$).

The evolution of both  R6 and R8 
flares in the $S_{m}$ -- $\nu_{m}$ plane is shown in Fig. \ref{radio_spectra} (c) -- (d). 
In the standard shock-in-jet model, $S_m \propto \nu_m ^{\epsilon_i}$ where $\epsilon_i$ depends upon 
the variation of physical quantities i.e., magnetic field (B), 
Doppler factor ($\delta$) and energy of relativistic electrons [see e.g. 1;3 for details]. 
The estimated $\epsilon_i$ values are given in 
Table \ref{fitted_para_SSA}.

\begin{table*}
\center
\caption{Different states of spectral evolution and their characteristics }
\begin{tabular}{c c c c c c c } \hline    
Flare  &Time          & bin     & $\epsilon_{Calculated}$ &  $\epsilon_{Expected}$               & b            &Stage      \\
       &JD [2454000+] &         &                         &    & s=2.2, a=1-2    &      \\\hline 
R6     &1096-1178     &1-4      & -7$\pm$3                &-2.5    & 0.7      & Compton    \\
       &1178-1194     &4-5      & 0                       &0       & -0.07    & Synchrotron \\
       &1194-1221     &5-8      & 10$\pm$2                &0.7     &2.6       & Adiabatic    \\
R8     &1283-1303     &13-15    & -0.9$\pm$0.1            &-2.5    &0.4       & Compton      \\
       &1298-1345     &15-18    & 1.8$\pm$0.2             &0.7     &-2        & Adiabatic     \\\hline 
\end{tabular} \\
$\delta \propto R^b$, $B \propto R^{-a}$ and $N(\gamma) \propto \gamma^{-s}$ \\
\label{fitted_para_SSA}
\end{table*}

We notice that there is a significant difference between the theoretically expected (from [1]) and our calculated $\epsilon$ 
values (see Table \ref{fitted_para_SSA}).  
Therefore, the rapid rise and decay of $S_m$ w.r.t. $\nu_m$ particularly in the case of the R6 flare  
(see Fig. \ref{radio_spectra}) rule out these simple assumptions of a constant Doppler factor ($\delta$). 
Consequently, we consider the evolution of radio flares including dependencies of physical parameters 
$a$, $s$ and $d$  following (7).  Here, $a$, $s$ and $d$ parametrize the variations 
of $\delta \propto R^b$, $B \propto R^{-a}$ and $N(\gamma) \propto \gamma^{-s}$ along the jet radius. 
Since it is evident that the $\epsilon$ values 
do not differ much for different choices of $a$ and $s$ [7], we assume for simplicity 
$s \approx$ constant. For the two extreme values of $a = 1$ and 2, we investigate the variations in $b$. 
The two different $a$ values give similar results for $b$.  
The calculated values of $b$ for the different stages of evolution of the radio flares are given in Table \ref{fitted_para_SSA}. 
{\it As a main result, we conclude that the Doppler factor varies significantly along the jet radius during the evolution 
of the two radio flares. }

\section{Correlated mm-gamma-ray variability}
\begin{figure}
\begin{minipage}[l]{0.5\linewidth}
\includegraphics[scale=0.35, angle=-90]{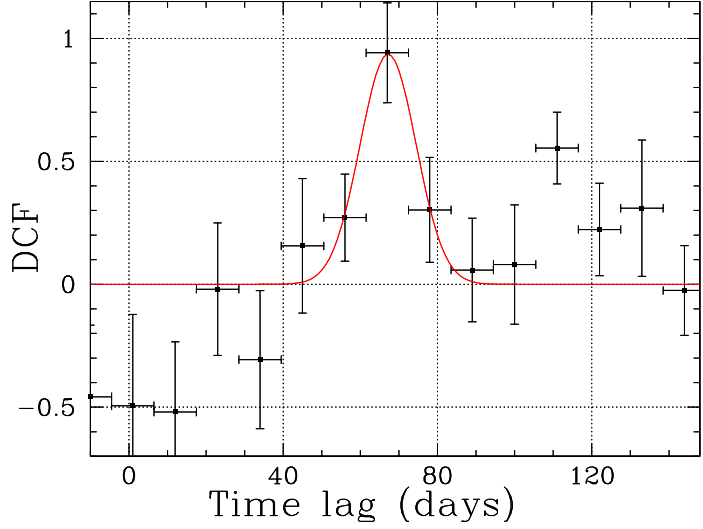}
\end{minipage}
\hspace{0.8in}
\begin{minipage}[r]{0.4\linewidth}
   \caption{Discrete cross-correlation function (DCF) of the $\gamma$-ray light curve w.r.t. the 230~GHz radio light 
curve.  The solid curve is the best 
fitted Gaussian function to the DCF curve binned at 11 days.               }
\end{minipage}
\label{dcf_gamma_230}
    \end{figure}

We apply the discrete cross-correlation function (DCF) [8] analysis method to investigate a 
possible correlation among flux variations at radio and 
$\gamma$-ray frequencies. In Fig. 3, we report the DCF analysis results of the weekly 
averaged $\gamma$-ray light curve with the 230~GHz radio data. 
To estimate the possible peak DCF value and respective time lag, 
we fit a Gaussian function to the DCF curve with a bin size of 11 days. The Gaussian function has a form: 
$DCF(t) = a \times ~exp [\frac{-(t-b)^{2}}{2c^{2}}]$, where $a$ is the peak value of 
the DCF, $b$ is the time lag at which the DCF peaks and $c$ characterizes the width of the Gaussian 
function.  The best-fit function is shown in 
Fig. \ref{dcf_gamma_230} and the fit parameters are $a = 0.94 \pm 0.30$, $b = (67 \pm 3)$~days and 
$c = (7 \pm 2)$~days. The significance of the correlation is checked using the linear Pearson correlation method which 
gives a confidence level $>$97$\%$. 
This indicates a clear correlation between the $\gamma$-ray and 230~GHz radio 
light curves of the source with the GeV flare leading the radio flare by $(67 \pm 3)$~days. 
Consequently, the flux variations at $\gamma$-rays lead those at radio frequencies $\sim$1 month time period, 
which suggests a non-cospatial origin of radio and $\gamma$-ray emission in the sense that $\gamma$-rays are produced closer to 
the central black hole.

\section{Summary and Conclusions} 
The evolution of the two major radio flares in the $\nu_m - S_m$ plane shows 
a very steep rise and decay over the Compton and adiabatic stages with a slope too steep to be explained from 
intrinsic variations, requiring an additional Doppler factor variation along the jet. 
For the two flares, we notice that $\delta$ changes as  $R^{0.7}$ during the rise and as $R^{2.6}$ during 
the decay of the R6 flare. The evolution of the R8 flare is governed by $\delta \propto R^{0.4}$ during the
rising phase and $\delta \propto R^{-2.0}$ during the decay phase of the flare.
Such a change in $\delta$ can be due to 
either a viewing angle ($\theta$) variation or a change of the bulk Lorentz factor ($\Gamma$) or by 
a combination of both. The change in $\delta$ can be easily interpreted as a few 
degree variation in $\theta$, while it requires a noticeable change of the bulk Lorentz factor. 
A similar behavior has also been observed in  
a parsec-scale VLBI kinematic study  of the source, which showed that the jet components 
exhibited significantly non-radial motion with regard to their position angle and in a direction perpendicular 
to the major axis of the jet [9]. Consequently, a correlation between the long-term radio flux-density variability 
and the position angle evolution of a jet component, implied  a significant geometric contribution 
to the origin of the long-term variability. 
This can be probably a result of precession at the base of the jet, which leads to twisted and/or helical structures. 
More observations and modeling is required to understand the physical origin of these phenomena. 
A formal cross-correlation between flux variations at radio and $\gamma$-ray frequencies 
suggests that $\gamma$-ray are produced closer to the black hole. The agreement of shock-induced evolution of 
radio flares with a clear correlation between radio and $\gamma$-rays is a hint for the shock-induced origin of 
$\gamma$-ray emission in the source. 
\\

\noindent 
{\bf Acknowledgments.}
The $Fermi$ LAT Collaboration acknowledges support from a number of agencies and institutes for both 
development and the operation of the LAT as well as scientific data analysis. These include NASA and DOE in 
the United States, CEA/Irfu and IN2P3/CNRS in France, ASI and INFN in Italy, MEXT, KEK, and JAXA in Japan, 
and the K.~A.~Wallenberg Foundation, the Swedish Research Council and the National Space Board in Sweden. 
Additional support from INAF in Italy and CNES in France for science analysis during the operations phase 
is also gratefully acknowledged. We would like to thank Marcello Giroletti and Stefanie Komossa for their 
useful comments and suggestions.

\providecommand{\href}[2]{#2}\begingroup\raggedright\endgroup


%

\end{document}